\documentclass[11pt]{article}
\usepackage{amsmath}
\usepackage{amsfonts}
\usepackage{amssymb}
\usepackage{theorem}
\usepackage{theorem}

\def\l{\lambda}
\def\p{\partial}
\def\e{\mathrm{e}}

\newcommand{\be}{\begin{equation}}
\newcommand{\ee}{\end{equation}}
\newcommand{\bea}{\begin{eqnarray}}
\newcommand{\eea}{\end{eqnarray}}
\newcommand{\beaa}{\begin{eqnarray*}}
\newcommand{\eeaa}{\end{eqnarray*}}

\newcommand{\nn}{\nonumber}
\renewcommand{\d}{\mathrm{d}}
\newcommand{\diverg}{\mathop{\mathrm{div}}\nolimits}
\begin{document}
\title{Dunajski-Tod equation and 
reductions of the generalized dispersionless 2DTL hierarchy}
\author{
L.V. Bogdanov\thanks
{L.D. Landau ITP RAS,
Moscow, Russia, e-mail
leonid@landau.ac.ru}}
\maketitle
\begin{abstract} We transfer the scheme 
for constructing differential reductions recently developed for the Manakov-Santini
hierarchy to the case of the two-component generalization of dispersionless 
2DTL hierarchy. We demonstrate that the equation arising as a result of the simplest
reduction is equivalent (up to a Legendre type transformation) 
to the Dunajski-Tod equation, locally describing
general ASD vacuum metric with conformal symmetry. We consider higher reductions
and corresponding reduced hierarchies also.
\end{abstract}
\section{Introduction}
In 1999 Dunajski and Tod \cite{DunTod99} introduced  the equation locally describing
general ASD vacuum metric with conformal symmetry, which reads 
\bea
(\eta F_{\tilde w}+F_{u\tilde w})(\eta F_{w}-F_{u w})-
(\eta^2 F-F_{u u})F_{w\tilde w}
=4e^{2\rho u}.
\label{DunTod99}
\eea
They also demonstrated that this equation is integrable, 
representing it as the integrability condition for a linear system of equations. 
More recently \cite{Dun08} Dunajski considered a case of ASD Ricci-flat metric 
with a conformal Killing vector whose self-dual derivative is null, and discovered
that the interpolating system describing this case is a simple differential reduction
of the Manakov-Santini system \cite{MS06,MS07}. This crucial observation initiated 
a study of reductions of the Manakov-Santini hierarchy \cite{LVB10MS}.

The Manakov-Santini system \cite{MS06,MS07} is a two-component generalization of
the dKP (Khohlov-Zabolotskaya) equation to the case of general vector fields in
the Lax pair, instead of Hamiltonian vector fields for the dKP case.  
If we don't impose Hamiltonian 
(area-preserving) reduction from the beginning, there is 
a freedom to consider more complicated reductions representing kind of twisted 
area-preservation conditions \cite{LVB10MS}. And it appeared that the reduction of the 
Manakov-Santini system corresponding to Dunajski interpolating system is the lowest 
order reduction of this class \cite{LVB10MS}. 
In terms of equations these reductions represent differential relations between dependent
variables, not reducing the dimension of the system 
(the number of independent variables). 
In this sense they are similar to the differential reductions known for the case of 
standard (`dispersionful') integrable systems (see, e.g., \cite{Zakharov1,BF}).

The Manakov-Santini hierarchy belongs
to the class of multidimensional hierarchies studied in \cite{BDM07,LVB09TMF}.
The construction of the work \cite{LVB10MS} can also be transferred 
to the multidimensional case \cite{LVB11}.

Similar to the Manakov-Santini system,
it is possible to introduce a two-component generalization of the dispersionless 
2DTL equation \cite{LVB10Toda} and consider its differential reductions. 
The general construction of interpolating reductions for the 
two-component generalization of the dispersionless 2DTL equation is developed 
in the present work.
The equation corresponding to the simplest reduction was presented in \cite{LVB10Toda}
in the form
\bea
m_{tt}=(m_t)^{\frac{1}{\alpha}}(m_{ty}m_x-m_{xy}m_t),
\label{intToda}
\eea
in the limit $\alpha\rightarrow 0$ it can be reduced to the d2DTL equation.
Taking into account that the equation (\ref{DunTod99}) for $\eta\rightarrow 0$
reduces  to the dispersionless 2DTL equation too \cite{Dun08}, 
it is a natural hypothesis that equations (\ref{DunTod99}) and (\ref{intToda}) 
should be connected, 
representing different forms of 
interpolating equation for the
d2DTL case. And it is indeed so! In the present paper we 
demonstrate that these equations are equivalent up to a Legendre type transformation.

First we introduce the two-component generalization of the dispersionless 2DTL 
equation and give an elementary description of interpolating reduction 
in terms of the Lax pair. We derive equation (\ref{intToda}) and demonstrate its
equivalence to Dunajski-Tod equation. We also discuss a connection of interpolating
equation (\ref{intToda}) for some special rational values of parameter $\alpha$ with
the generalized dispersionless Harry Dym equation constructed by Blaszak 
\cite{Blaszak02PLA,Blaszak02JPA}. 
Finally, we develop a general construction of differential (interpolating)
reductions of the generalized dispersionless 2DTL hierarchy and present 
the hierarchy connected with equation (\ref{intToda}).
\section{Interpolating reduction: elementary description}
First we will give a description of the simplest interpolating reduction
of the two-component generalization of the dispersionless 2DTL 
equation in terms of the Lax pair. A
general construction of reductions in terms of the hierarchy is presented below.
\subsection{Two-component generalization of the dispersionless 2DTL equation}
The simplest two-component generalization of the dispersionless 2DTL equation 
was introduced in \cite{LVB10Toda} as a first equation 
of the hierarchy (see Section \ref{Hierarchy})
generalizing dispersionless 2DTL hierarchy to the case of non-Hamiltonian vector
fields, it reads
\bea
&&
({e}^{-\phi})_{tt}=m_t\phi_{xy}-m_x\phi_{ty},
\nn\\
&&
m_{tt}{e}^{-\phi}=m_{ty}m_x-m_{xy}m_t,
\label{gen2DTL}
\eea
and the corresponding Lax pair is
\bea
&&
\partial_x\mathbf{\Psi}=\left((\l+ \frac{m_x}{m_t})\partial_t - 
(\phi_t \frac{m_x}{m_t}-\phi_x)\l\partial_\l\right)\mathbf{\Psi},
\nn\\
&&
\partial_y\mathbf{\Psi}=\left(-\frac{1}{\l}\frac{{e}^{-\phi}}{m_t}\partial_t -
\frac{1}{\l}\frac{({e}^{-\phi})_t}{m_t}\l\partial_\l\right)\mathbf{\Psi}.
\label{Laxgen2DTL}
\eea
For $m=t$ the system (\ref{gen2DTL}) reduces to the dispersionless 2DTL equation
\bea
({e}^{-\phi})_{tt}=\phi_{xy},
\label{d2DTL}
\eea
Respectively, the reduction $\phi=0$ gives an equation 
\cite{Pavlov03} (see also \cite{MASh02}, \cite{MASh04})
\bea
m_{tt}=m_{ty}m_x-m_{xy}m_t.
\label{Spa}
\eea 
\subsection{Interpolating reduction}
A standard way to define a reduction starting from the Lax pair 
is to suggest that Lax equations possess a solution
$f$ with some special analytic properties in $\lambda$ invariant under dynamics
(i.e., belonging to some invariant manifold of linear equations). Suggesting that
that Lax equations (\ref{Laxgen2DTL}) possess a solution polinomial in $\lambda$
or $\lambda^{-1}$, we arrive to Gelfand-Dikii reductions of the two-component
d2DTL (\ref{gen2DTL}), which for d2DTL equation (\ref{d2DTL}) reduce to standard
Gelfand-Dikii reductions. A simplest case of polynomial solution $f=\lambda$
leads to equation (\ref{Spa}), and d2DTL equation just degenerates in this case.

Interpolating reductions of the Manakov-Santini hierarchy 
introduced in \cite{LVB10MS}
generalize Gelfand-Dikii reductions, and, containing a parameter, 'interpolate'
between the Manakov-Santini hierarchy 
Gelfand-Dikii reduction of the order $n$ and the dKP hierarchy.
To understand the origin of this type of reductions in terms of the Lax pair 
for the system (\ref{gen2DTL}), it is important to note that there exists a whole
one-parametric family of Lax pairs (or overdetermined systems) corresponding 
to the system (\ref{gen2DTL}), of which the Lax pair (\ref{Laxgen2DTL}) is only
a special representative. It is rather general fact
for the Lax pairs in terms of
non-Hamiltonian or not divergence-free vector fields.

Indeed, starting from the Lax pair (\ref{Laxgen2DTL}), written symbolically in the
form, 
\bea
&&
\partial_x {\Psi}=\hat U\Psi,
\nn\\
&&
\partial_y {\Psi}=\hat V\Psi,
\label{UV1}
\eea
where $\hat U=u_1\p_t+u_2\l\p_\l$, $\hat V=v_1\p_t+v_2\l\p_\l$ are vector
fields defined by (\ref{Laxgen2DTL}), 
we inroduce a one-parametric family of Lax pairs
\bea
&&
\partial_x {\Phi}=\hat U\Phi +\beta \diverg \hat U\Phi,
\nn\\
&&
\partial_y {\Phi}=\hat V\Phi +\beta \diverg \hat V\Phi,
\label{UV2}
\eea
where $\beta$ is a parameter, $\diverg \hat U=\p_t u_1+\l\p_\l u_2$, 
$\diverg \hat V=\p_t v_1+\l\p_\l v_2$. These Lax pairs are no more in the form
of pure vector fields, they also have a non-differential term. However, it is
easy to check that compatibility conditions for them remain the same and provide system
(\ref{gen2DTL}). It is convenient to rewrite the system (\ref{UV2}) 
in terms of $\ln \Phi$,
then it takes the form of nonhomogeneous linear system
\bea
&&
\partial_x \ln{\Phi}=\hat U\ln \Phi +\beta \diverg \hat U,
\nn\\
&&
\partial_y \ln {\Phi}=\hat V \ln \Phi +\beta \diverg \hat V,
\label{UV3}
\eea

Knowing the general solution of the linear system
(\ref{UV1}), it is not difficult to construct the general solution for the systems
(\ref{UV2}), (\ref{UV3}). 
Indeed, let $\Psi_1$, $\Psi_2$ be two solutions of system (\ref{UV1})
with nonzero Jacobian (Poisson bracket) $J=\{\Psi_1,\Psi_2\}\neq 0$, 
$\{f,g\}=\l(f_\l g_t-f_t g_\l)$.
Then the general solution of the system (\ref{UV1}) is of the form 
$F(\Psi_1,\Psi_2)$. The Poisson bracket of two solutions satisfies the
system 
\bea
&&
\partial_x \ln {J}=\hat U \ln J + \diverg \hat U ,
\nn\\
&&
\partial_y \ln {J}=\hat V \ln J +\diverg \hat V,
\label{UV20}
\eea
having a general solution $\ln\{\Psi_1,\Psi_2\} + f(\Psi_1,\Psi_2)$,
where the first term is a special solution of the nonhomogeneous equations,
and the second term is a general solution of the homogeneous equations.
Compairing this system to the system (\ref{UV3}), we conclude that
the general solution of the system (\ref{UV3}) is
$\ln\Phi=\beta \ln\{\Psi_1,\Psi_2\} + f(\Psi_1,\Psi_2)$,
and the general solution of the system (\ref{UV2}) is 
\beaa
\Phi=\{\Psi_1,\Psi_2\}^{\beta} F(\Psi_1,\Psi_2).
\eeaa

Suggesting the existence of solution 
$f$ with some special analytic properties in $\lambda$
for the Lax pairs (\ref{UV2}) or (\ref{UV3}), 
we will obtain one-parametric
interpolating reduction, which for $\beta=0$ implies the existence of solution
$f$ for standard Lax equations (\ref{Laxgen2DTL}) 
(Gelfand-Dikii type reduction), and in the
limit $\beta\rightarrow\infty$ corresponds to Hamiltonian (divergence-free)
vector fields.

We define a simplest interpolating reduction for equation (\ref{gen2DTL})
by the condition that Lax equations (\ref{UV2}) possess a solution
$f=\lambda$ (equivalently, equations (\ref{UV3}) possess a solution $\ln \l$
and equations (\ref{UV20}) -- solution $-\alpha\ln \l$, $\alpha=-\beta^{-1}$.)
An explicit form of equations (\ref{UV3}) corresponding to (\ref{Laxgen2DTL})
is 
\bea
&&
\partial_x\ln{\Phi}=\left((\l+ \frac{m_x}{m_t})\partial_t - 
(\phi_t \frac{m_x}{m_t}-\phi_x)\l\partial_\l\right)\ln{\Phi}
+\beta \partial_t \frac{m_x}{m_t},
\nn\\
&&
\partial_y\ln{\Phi}=\left(-\frac{1}{\l}\frac{{e}^{-\phi}}{m_t}\partial_t -
\frac{1}{\l}\frac{({e}^{-\phi})_t}{m_t}\l\partial_\l\right)\ln{\Phi}-
\beta\frac{{e}^{-\phi}}{\l}
\partial_t\frac{1}{m_t},
\label{Laxgen2DTLbeta}
\eea
and it is easy to check that substitution of solution $\ln \l$
($\Phi=\lambda$) to both
equations gives the same reduction condition
\bea
\e^{\alpha\phi}=m_t, \quad \alpha=-\beta^{-1},
\label{intreduction}
\eea
which is an interpolating reduction of the system (\ref{gen2DTL}).
This reduction makes it possible to rewrite the system (\ref{gen2DTL}) as
one equation for $m$ (\ref{intToda})
or in the form of deformed d2DTL equation,
\beaa
&&
({e}^{-\phi})_{tt}=m_t\phi_{xy}-m_x\phi_{ty},
\nn\\
&&
m_t=\e^{\alpha\phi}.
\eeaa
The limit $\alpha\rightarrow 0$ corresponds to the dispersionless 2DTL equation
(\ref{d2DTL}), and the limit $\alpha\rightarrow\infty$ gives equation (\ref{Spa}).

\subsection{Legendre transform and Dunajski-Tod interpolating equation}
Equation (\ref{intToda}) can be represented in exterior differential form 
\bea
\gamma^{-1}\d m_t^\gamma\wedge \d x \wedge \d y=\d m_y\wedge\d m\wedge\d y,
\label{extint}
\eea
where $\gamma=1-\alpha^{-1}$.

Let us consider a
Legendre type transform (where $\tau$ is a new independent variable and $M$
is a new dependent variable)
\beaa
m_t=e^\tau,\quad M=m-te^\tau.
\eeaa
This transform is suggested to be non-degenerate (at least locally), 
some special cases and global behavior may require more accurate analysis.
Differential of $M$ is of the form
\beaa
\d M=M_x\d x+ M_y \d y - te^\tau \d\tau.
\eeaa
Transformed equation (\ref{extint}) reads
\beaa
\gamma^{-1}\d e^{\gamma\tau}\wedge \d x \wedge \d y=\d M_y\wedge\d M\wedge\d y-
\d M_y\wedge\d M_\tau\wedge\d y,
\eeaa
and transformed equation (\ref{intToda}) looks like
\bea
e^{\gamma\tau}=(M_{y\tau}M_x-M_{yx}M_\tau)-(M_{y\tau}M_{x\tau}-M_{yx}M_{\tau\tau}).
\label{intToda1}
\eea
Scaling the time $\tau\rightarrow 2\tau$, we get
\beaa
4e^{2\gamma\tau}=2(M_{y\tau}M_x-M_{yx}M_\tau)-(M_{y\tau}M_{x\tau}-M_{yx}M_{\tau\tau})
\eeaa
In terms of  the function $F=e^{-\tau} M$
\beaa
(F_{y}+F_{y\tau})(F_{x}-F_{x\tau})-(F-F_{\tau\tau})F_{xy}=4e^{-2\alpha^{-1}\tau}.
\eeaa
Considering the scaling  $x\rightarrow \eta^{-1}x$, $y\rightarrow \eta^{-1}y$,
$\tau\rightarrow \eta \tau$, we obtain Dunajski-Tod equation
\bea
(\eta F_{y}+F_{y\tau})(\eta F_{x}-F_{x\tau})-(\eta^2 F-F_{\tau\tau})F_{xy}
=4e^{2\rho\tau},
\eea
where $\rho=-\alpha^{-1}\eta$.
\subsection{Generalized dispersionless Harry Dym equation and d2DTL
interpolating equation}
Generalized dispersionless Harry Dym equation constructed by Blaszak 
\cite{Blaszak02PLA,Blaszak02JPA} 
can be written in the form of conservation law,
\bea
\p_t u^{2-r}=\frac{(3-r)}{(r-1)(r-2)}
\left(
u^{2-r} \p_x^{-1} \p_y u^{r-1}
\right)_y,
\label{HD}
\eea
where parameter $r$ is integer, $r\in \mathbb{Z}$. This equation suggests
the existence of potential $v$, such that
\beaa
&&
\p_y v=u^{2-r},\\
&&
\p_t v=\frac{(3-r)}{(r-1)(r-2)}u^{2-r} \p_x^{-1} \p_y u^{r-1},
\eeaa
and for the potential we get an equation
\beaa
\frac{(3-r)}{(r-1)(r-2)} v_{yy}=v_y^{\frac{r-3}{2-r}}(v_{xt}v_y-v_t v_{xy}),
\eeaa
which after the change of variables $y\rightarrow t$, $x\rightarrow y$,
$t\rightarrow x$ is equivalent to equation (\ref{intToda}) with 
$\alpha=\frac{2-r}{r-3}$ (the constant can be eliminated by the rescaling
of variables).
Thus the generalized dispersionless Harry Dym equation is connected with equation
(\ref{intToda}) with some rational values of parameter $\alpha$. The special values
$\alpha=\infty$ corresponding to equation (\ref{Spa}) and $\alpha=0$ corresponding
to d2DTL are related respectively with special values $r=3$ and $r=2$ for the
generalized dispersionless Harry Dym equation (\ref{HD}). 

Several examples
of integrable equations with a structure similar to equations (\ref{HD}), 
(\ref{intToda}) were provided in the work \cite{Fer10}.
\section{Interpolating reductions: general construction}
\label{Hierarchy}
\subsection{Generalized dispersionless 2DTL hierarchy} 
First we briefly describe the generalized dispersionless 2DTL hierarchy,
following the work \cite{LVB10Toda} (on d2DTL hierarchy see
\cite{TT91}, \cite{TT95}, \cite{MS09}).
 
We consider formal series
\bea
&&
\Lambda^\text{out}=\ln\lambda+\sum_{k=1}^{\infty}l^+_k\lambda^{-k},\quad
\Lambda^\text{in}=\ln\lambda+\phi+\sum_{k=1}^{\infty}l^-_k\lambda^{k},
\nn
\\
&&
M^\text{out}=M_0^\text{out} + \sum_{k=1}^{\infty}m^+_k\e^{-k\Lambda^\text{out}},\quad
M^\text{in}=M_0^\text{in} + m_0 +\sum_{k=1}^{\infty}m^-_k\e^{k\Lambda^\text{in}},
\nn
\\
&&
M_0=t+\sum_{k=1}^{\infty}x_k\e^{k\Lambda}-
\sum_{k=1}^{\infty}y_k\e^{-k\Lambda},
\label{LMseries}
\eea
where  $\lambda$ is a spectral variable, $t$,  
$x_k$, $y_k$ are considered independent
variables, and other coefficients of the series 
($\phi$, $m_0$, $l^{\pm}_k$, $m^{\pm}_k$) -- dependent variables.
Usually we
suggest that `out' and `in' components of the series define the functions
outside and inside the unit circle in the complex plane of the variable
$\lambda$ respectively (in more detail in \cite{LVB10Toda}). 
Generalized dispersionless 2DTL hierarchy is defined by the generating relation
\be
(\{\Lambda,M\}^{-1}\d\Lambda\wedge \d M)^\text{out}=
(\{\Lambda,M\}^{-1}\d\Lambda\wedge \d M)^\text{in},
\label{genToda}
\ee
where $\{f,g\}=\l(f_\l g_t-f_t g_\l)$,
which may be considered as a continuity condition on the unit circle for the differential
two-form (or just in terms of formal series);
$\{\Lambda,M\}^\text{out}=1+O(\lambda^{-1})$, 
$\{\Lambda,M\}^\text{in}=1+\p_t m_0 + O(\lambda)$,
and we suggest that $\{\Lambda,M\}\neq 0$.
The differential $\d$ is given by 
\be
\d f=\p_\lambda f \d\lambda + \p_t f\d t+
\sum_{k=1}^{\infty}\partial^+_k f \d x_k+
\sum_{k=1}^{\infty}\partial^-_k f\d y_k,
\label{diff}
\ee
where $\partial^+_k f=\frac{\p f}{\p x_k}$, $\partial^-_k f=\frac{\p f}{\p y_k}$.
As a result of condition (\ref{genToda}), 
the coefficients of the differential two-form in the generating relation
(\ref{genToda}) are {\em meromorphic}.

Generating equation (\ref{genToda}) implies Lax-Sato equations of the hierarchy.
In explicit form, a complete set of Lax-Sato equations reads
\bea
&&
\left(
\partial^+_n
-\left(\frac{\e^{n\Lambda}\l\p_\l\Lambda}
{\{\Lambda,M\}}
\right)^\text{out}_+\p_t
+ 
\left(\frac{\e^{n\Lambda}\p_t\Lambda}
{\{\Lambda,M\}}
\right)^\text{out}_+
\l\p_\l
\right)
\begin{pmatrix}
\Lambda\\
M
\end{pmatrix}=0,\qquad
\label{Hi1}
\\
&&
\left(
\partial^-_n
+\left(\frac{\e^{-n\Lambda}\l\p_\l \Lambda}
{\{\Lambda,M\}}
\right)^\text{in}_-\p_t
- 
\left(\frac{\e^{-n\Lambda}\p_t\Lambda}
{\{\Lambda,M\}}
\right)^\text{in}_-
\l\p_\l
\right)
\begin{pmatrix}
\Lambda\\
M
\end{pmatrix}=0,\qquad
\label{Hi2}
\eea
where $(\dots)_-$, $(\dots)_+$ are standard projections respectively to negative
and nonnegative powers of $\lambda$.
The Lax-Sato equations for the times $x=x_1$, $y=y_1$,
$\partial^+_1=\p_x$, $\partial^-_1=\p_y$,
\bea
&&
\partial_x\mathbf{\Psi}=\left((\l+ (m_1^+)_t-l_1^+)\partial_t - 
\l (l_1^+)_t\partial_\l\right)\mathbf{\Psi},
\label{firstLaxSato0}
\\
&&
\partial_y\mathbf{\Psi}=\left(-\frac{1}{\l}\frac{\mathrm{e}^{-\phi}}{m_t}\partial_t -
\frac{(\mathrm{e}^{-\phi})_t}{m_t}\partial_\l\right)\mathbf{\Psi},
\label{firstLaxSato}
\eea
where $\mathbf{\Psi}=\begin{pmatrix}
\Lambda\\
M
\end{pmatrix}$, $m=m_0+t$,
correspond to the Lax pair (\ref{Laxgen2DTL}), where the coefficients 
in the first Lax-Sato equation can be transformed to the form (\ref{Laxgen2DTL})
by taking its expansion at $\lambda=0$, and the system (\ref{gen2DTL}) arises as
a compatibility condition.

Lax-Sato equations (\ref{Hi1},\ref{Hi2}) define the evolution of the series
$\Lambda^\text{in},\Lambda^\text{out}$, $M^\text{in},M^\text{out}$. The only term
containing an interaction between $\Lambda$ and $M$ is $\{\Lambda,M\}$.
The condition $\{\Lambda,M\}=1$ splits out equations for $\Lambda$ and
reduces the hierarchy
(\ref{Hi1},\ref{Hi2}) to the d2DTL hierarchy, while the condition $\Lambda=\ln \l$ --
to the hierarchy, considered by Mart\'{i}nez Alonso and Shabat
\cite{MASh02,MASh04}, see also Pavlov \cite{Pavlov03}.
\subsection{Interpolating reduction for the hierarchy}
Following the scheme of the work \cite{LVB10MS}, we start from nonhomogeneous
Lax-Sato equations for the Jacobian (Poisson bracket) $J_0=\{\Lambda,M\}$.
We define reductions by the condition that some specific solution of these
equations is continuous on the unit circle ('in' and 'out' components are
equal). This condition is preserved by
the dynamics because the coefficients of vector fields in the Lax-Sato equations
are meromorphic with respect to $\lambda$, and thus it defines a reduction
of the hierarchy.

Rewriting Lax-Sato equations (\ref{Hi1},\ref{Hi2})  
symbolically as (compare (\ref{UV1}))
\bea
({\partial^+_n}
-\hat U_n)
\begin{pmatrix}
\Lambda\\
M
\end{pmatrix}=0,
\qquad
({\partial^-_n}
-\hat V_n)
\begin{pmatrix}
\Lambda\\
M
\end{pmatrix}=0,
\label{UVSato}
\eea
where corresponding vector fields are defined explicitly by Lax-Sato equations
(\ref{Hi1},\ref{Hi2}),
we obtain nonhomogeneous linear
equations for the Jacobian in the form (see also (\ref{UV20}))
\bea
&&
\partial^+_n \ln {J_0}=\hat U_n \ln J_0 + \diverg \hat U_n ,
\nn\\
&&
\partial^-_n \ln {J_0}=\hat V_n \ln J_0 +\diverg \hat V_n.
\label{UV21}
\eea

We define interpolating reduction for the hierarchy by the condition
\bea
(\ln J_0-\alpha \Lambda)^\text{out}=(\ln J_0-\alpha \Lambda)^\text{in}
\label{Int0}
\eea
Both sides of this relation satisfy equations (\ref{UV21}), which preserve
the continuity, and thus condition (\ref{Int0}) indeed defines a reduction.

This relation implies that 
\bea
(\ln J_0-\alpha \Lambda)=-\alpha \ln \lambda,  
\label{Jlambda}
\eea
thus nonhomogeneous linear equations of the hierarchy
(\ref{UV21}) possess a solution $f=-\alpha \ln \lambda$. This property was
used above to define interpolating reduction in terms of Lax pair for the system
(\ref{intToda}). It is possible to obtain reduction condition (\ref{intreduction})
directly from relation (\ref{Jlambda}), taking its expansion at $\lambda=0$,
where at order zero we get 
\bea
{\alpha\phi}=\ln(1+\p_t m_0)=\ln m_t.
\label{intredHi}
\eea

Substituting the expression for the Poisson bracket implied by relation
(\ref{Jlambda}),
\beaa
J_0=\{\Lambda,M\}=\lambda^{-\alpha}\exp(\alpha \Lambda),
\eeaa
to the generating relation (\ref{genToda}),
we obtain the generating relation for the reduced hierarchy
\beaa
(\exp(-\alpha \Lambda)\d\Lambda\wedge\d M)^\text{out}=
(\exp(-\alpha \Lambda)\d\Lambda\wedge\d M)^\text{in}.
\eeaa
The two-form $\Omega$ defined by the generating relation,
\beaa
\Omega=(\d\Lambda\wedge\d(\exp(-\alpha \Lambda) M))^\text{out}=
(\d\Lambda\wedge\d(\exp(-\alpha \Lambda) M))^\text{in},
\eeaa
is (up to a factor $\lambda^{-\alpha}$)
meromorphic in the complex plane having poles only at zero and infinity,
and it is evidently closed. The condition of conservation of this form 
can be used to define a reduction in terms of nonlinear vector Riemann-Hilbert
problem and develop a dressing scheme for the reduced hierarchy similar to 
\cite{LVB10MS}
(see also \cite{LVB10Toda}).

The Lax-Sato equations for the reduced hierarchy read
\beaa
&&
\left(
\partial^+_n
-\left({\lambda^\alpha\e^{(n-\alpha)\Lambda}\l\p_\l\Lambda}
\right)^\text{out}_+\p_t
+ 
\left({\lambda^\alpha\e^{(n-\alpha)\Lambda}\p_t\Lambda}
\right)^\text{out}_+
\l\p_\l
\right)
\begin{pmatrix}
\Lambda\\
M
\end{pmatrix}=0,\qquad
\nn
\\
&&
\left(
\partial^-_n
+\left(\lambda^\alpha{\e^{(-n-\alpha)\Lambda}\l\p_\l \Lambda}
\right)^\text{in}_-\p_t
- 
\left(\lambda^\alpha{\e^{(-n-\alpha)\Lambda}\p_t\Lambda}
\right)^\text{in}_-
\l\p_\l
\right)
\begin{pmatrix}
\Lambda\\
M
\end{pmatrix}=0.\qquad
\nn
\eeaa
Similar to d2DTL hierarchy, Lax-Sato equations for $\Lambda$ split out,
having no interaction with $M$.
\subsection{Higher reductions}
Here we define a class of reductions interpolating between Gelfand-Dikii reductions
of the order $n$ for the generalized d2DTL hierarchy and d2DTL hierarchy proper.
Gelfand-Dikii reductions suggest existence of rational solution for 
Lax-Sato equations (\ref{Hi1}), (\ref{Hi2}).
In the case of d2DTL hierarchy Gelfand-Dikii reduction 
of the order $n$ implies stationarity with respect to
a higher flow
$\p_n=a\p^+_n+b\p^-_n$, where $a$, $b$ are some constants. However,
for the case of non-Hamiltonian vector fields this is not true.

We define higher interpolating reductions by the condition
\bea
(\ln J_0 + a L^n + b L^{-n})^\text{out}=(\ln J_0 + a L^n + b L^{-n})^\text{in},
\label{higherred}
\eea
where $L=e^\Lambda$. Both sides of this condition are solutions of nonhomogeneous
Lax-Sato equations (\ref{UV21}), and due to continuity 
they are equal to a single rational function 
$f=a (L^n)^\text{out}_+ + b (L^{-n})^\text{in}_-$.
Thus the reduction condition (\ref{higherred}) implies that nonhomogeneous
Lax-Sato equations (\ref{UV21}) possess a rational solution $f$. This property
can be used to define a reduction in terms of the Lax pair for the system
(\ref{gen2DTL}) and calculate a differential reduction in terms of the functions
$\phi$, $m$. In the limit $a,\,b\rightarrow 0$ the reduction leads to $J_0=1$
and thus to d2DTL hierarchy, the limit $a,\,b\rightarrow \infty$ implies that
homogeneous Lax-Sato equations (\ref{Hi1}), (\ref{Hi2}) possess a rational
solution $f$, and thus it is a Gelfand-Dikii reduction of the order $n$.

Reduction condition (\ref{higherred}) implies the expression for the Poisson
bracket,
\bea
J_0=\{\Lambda,M\}=\exp(a (L^n)^\text{out}_+ +b (L^{-n})^\text{in}_-
-a L^n - bL^{-n}),
\label{Jn}
\eea
which is valid for both 'in' and 'out' components. Substituting this expression
to the generating relation (\ref{genToda}), we obtain the generating relation for 
the reduced hierarchy
\beaa
&&
(\exp(a L^n + bL^{-n}-a (L^n)^\text{out}_+ 
-b (L^{-n})^\text{in}_-)\d\Lambda\wedge\d M)^\text{out}
\\
&&\qquad
=(\exp(a L^n + bL^{-n} - a (L^n)^\text{out}_+ 
- b (L^{-n})^\text{in}_-)\d\Lambda\wedge\d M)^\text{in}.
\eeaa
We will not write down explicitly the Lax-Sato equations for the reduced
hierarchy, it is rather straightforward substituting
(\ref{Jn}) to (\ref{Hi1}), (\ref{Hi2}). A characteristic feature of these equations
is that due to the reduction the equations for $\Lambda$ split out, similar
to d2DTL hierarchy.

Let us consider a case $n=1$ in more detail. Nonhomogeneous
Lax-Sato equations (\ref{UV21}) in this case possess a rational solution
\beaa
f=a L^\text{out}_+ + b (L^{-1})^\text{in}_-
=a(\lambda+l^+_1) + b \frac{e^{-\phi}}{\lambda}
\eeaa
(we use the series  (\ref{LMseries}), $L=e^\Lambda$). There are two strategies
to  calculate the differential reduction for the system
(\ref{gen2DTL}) in terms of the functions
$\phi$, $m$ (which can be used for reductions of arbitrary order $n$).
One is to substitute the rational function $f$ to  
nonhomogeneous Lax pair (\ref{Laxgen2DTLbeta}). The arising conditions define all
coefficients of the function and give a reduction condition.
Another way is to take an expansion of relation (\ref{Jn}) and
use the Lax pair (\ref{Laxgen2DTL}) 
(see also (\ref{firstLaxSato0},\,\ref{firstLaxSato})) 
to express higher coefficients of the series for $\Lambda$, $M$ through
$\phi$, $m$ (in general in the form of recursion relations).
In the case $n=1$ the zero order of expansion of relation (\ref{Jn})
at $\lambda=0$ gives
\beaa
\ln m_t=al^+_1 + be^{-\phi}l_1^-.
\eeaa
Using the pair (\ref{Laxgen2DTL}), we obtain following 
relations for $l^+_1$, $l_1^-$:
\beaa
&&
\p_t l^+_1=\phi_t \frac{m_x}{m_t}-\phi_x,
\quad
\p_y l^+_1=-\frac{(e^{-\phi})_t}{m_t},
\\
&&
\p_t (e^{-\phi}l^-_1)=-m_t\phi_y,
\quad 
\left(\frac{m_x}{m_t}\p_t - \p_x\right)(e^{-\phi}l^-_1)=(e^{-\phi})_t.
\eeaa
The simplest form of the differential reduction for the system (\ref{gen2DTL})
in the case $n=1$ reads
\beaa
m_{tt}=a(\phi_t {m_x}- \phi_x {m_t}) - b(m_t)^2\phi_y.
\eeaa
However, in this form the differential relation contains all the variables
$x$, $y$, $t$.
It is possible rewrite it in equivalent form in $(x,t)$
plane or $(y,t)$ plane. The differential reduction in $(y,t)$ plane reads
\beaa
\p_y\p_t \ln m_t= -a\p_t \left( \frac{(e^{-\phi})_t}{m_t}\right)
-b\p_y (m_t\phi_y).
\eeaa
Considering this reduction together with interpolating reduction
(\ref{intreduction}) (see also (\ref{intredHi}))
\beaa
\e^{\alpha\phi}=m_t, 
\eeaa
we obtain a (1+1)-dimensional system which represents a reduction of
equation (\ref{intToda}) and can be rewritten as (1+1)-dimensional equation
for the function $m$
\bea
am_{tt}=(m_t)^{{\frac{1}{\alpha}}+1}(\alpha m_{ty} + b m_{yy}m_t).
\label{(1+1)a}
\eea
It is possible to transform this equation to the system of hydrodynamic type.

The differential reduction in $(x,t)$ plane reads
\beaa
\p_t \left( a\frac{m_x}{m_t}(\phi_t \frac{m_x}{m_t}-\phi_x) +
\p_t\frac{m_x}{m_t} + b (e^{-\phi})_t
\right)
- a\p_x \left(\phi_t \frac{m_x}{m_t}-\phi_x\right)=0.
\eeaa
Considered together with interpolating reduction
(\ref{intreduction}), it forms a (1+1)-dimensional system representing
a reduction of equation (\ref{intToda}), which in terms of one function
$m$ reads
\bea
bm_{tt}m_t^{-\frac{1}{\alpha}-1}-\alpha\p_t \frac{m_x}{m_t}
+ a\left(\frac{m_x}{m_t}\p_t-\p_x\right) \frac{m_x}{m_t}=0.
\label{(1+1)b}
\eea
A common solution of (1+1)-dimensional equations (\ref{(1+1)a}), (\ref{(1+1)b})
gives a solution of equation (\ref{intToda}).
\section*{Acknowledgements}
The author is grateful to M. Dunajski
for suggesting that equations (\ref{DunTod99}), (\ref{intToda})
could be connected and useful discussions. The author would like to
thank M. Pavlov for demonstrating the relation between the 
generalized dispersionless Harry Dym equation introduced by Blaszak
(\ref{HD}) and equation (\ref{intToda}). The author also acknowledges interesting
discussions with S.V. Manakov and P.M. Santini on the topics connected with
this work. This research
was partially supported by the Russian Foundation for Basic Research under grant
no 10-01-00787 and by the President of Russia grant 6170.2012.2 (scientific schools).

\end{document}